\documentclass[10pt,a4paper]{article}
\usepackage{epsf}
\usepackage{graphicx}
\usepackage{natbib}
\usepackage{amsmath,amssymb}
\usepackage{url}
\usepackage{cite}
\usepackage{xcolor}

\usepackage{journals_joge}

\def\fracd#1#2{\frac{\displaystyle #1}{\displaystyle #2}}
\newcommand{\iu}{{\mathrm{i}\mkern1mu}}
\newcommand{\tX}{\mathsf{X}}
\newcommand{\tY}{\mathsf{Y}}
\newcommand{\bfS}{\mathbf{S}}
\newcommand{\bfX}{\mathbf{X}}
\newcommand{\bfY}{\mathbf{Y}}
\newcommand{\rmH}{\mathrm{H}}
\newcommand{\rmT}{\mathrm{T}}

\begin{document}

\title{Filling the gap in the IERS C01 polar motion series in 1858.9--1860.9}
\author{{\Large Zinovy Malkin$^1$, Nina Golyandina$^2$, Roman Olenev$^3$}\\[1em]
  $^1$Pulkovo Observatory, St.~Petersburg, 196140, Russia, malkin@gaoran.ru\\
  $^2$Faculty of Mathematics and Mechanics, St.~Petersburg State University,\\
  St.~Petersburg, 199034, Russia, n.golyandina@spbu.ru\\
  $^3$Faculty of Mathematics and Mechanics, St.~Petersburg State University,\\
  St.~Petersburg, 199034, Russia, roman.olenev.cs@mail.ru}
\date{May 28, 2025}
\maketitle

\begin{abstract}
The C01 Earth orientation parameters (EOP) series provided by the
International Earth Rotation and Reference Systems Service (IERS)
is the longest reliable record of the Earth rotation.
In particular, the polar motion (PM) series beginning from 1846
provides a basis for investigation of the long-term PM variations.
However, the pole coordinate $Y_p$ in the IERS C01 PM series has
a 2-year gap, which makes this series not completely evenly spaced.
This paper presents the results of the first attempt to overcome
this problem and discusses some ways to fill this gap.
Two novel approaches were considered for this purpose:
parametric astronomical model consisting of the bias and the Chandler
and annual wobbles with linearly changing amplitudes, and
data-driven model based on Singular Spectrum Analysis (SSA).
Both methods were tested with various options to ensure robust
and reliable results.
The results obtained by the two methods generally agree within the $Y_p$
errors in the IERS C01 series, but the results obtained by the SSA
approach can be considered preferable because it is based on a more
complete PM model.
\end{abstract}

%%%%%%%%%%%%%%%%%%%%%%%%%%%%%%%%%%%%%%%%%%%%%%%%%%%%%%%%%%%%%%%%%%%%%%%%%%%%%%%%%%%%%%%%%%%%%%

\section{Introduction}

Time series of the Earth orientation parameters (EOP)
are the primary source of information for investigation
of the Earth rotation, in particular the polar motion (PM).
PM observational data are reported by Earth rotation services as
the coordinates $X_p$ and $Y_p$ of the Celestial Intermediate Pole (CIP)
in the International Terrestrial Reference System (ITRS).
The theoretical foundations of the PM theory are explained, e.~g.,
in \citet{Gross1992GJI,IERSConv2010} and papers referenced therein.
Historical overview of the PM observations and analyses
can be found in \citet{IAUC178}.
The PM is a complex phenomenon, which consists
of many variational components including trends and
(quasi)periodical oscillations with periods from several
hours to several decades.
In particular, the study of long-term polar motion variations is crucial
for expanding our understanding of Earth's dynamic processes, particularly
in terms of the impacts of climate change, changes in Earth's internal
structure and the redistribution of global mass.
To study slow PM variations, long time series of
the pole coordinates are needed.
The longer the time interval covered by EOP series,
the lower frequency variations can be detected and
investigated.

The international standard of the EOP series is the combined
series computed by the International Earth Rotation
and Reference Systems Service (IERS) \citep{Bizouard2019}.
The IERS C01 EOP series contains the longest PM series derived
from astronomical observations lasting from 1846 to
now\footnote{\url{ftp://hpiers.obspm.fr/iers/eop/eopc01/eopc01.iau2000.1846-now}}.
It is widely used for analysis of the decadal variations
in Earth rotation
\citep{Gaposchkin1972,Vondrak1988IAUS,Rykhlova1990IAUS,Kolaczek1999srst,Gambis2000ASPC,%
Yatskiv2000ASPC,Schuh2001JGeod,Hopfner2004SGeo,Guo2005JGeod,Malkin2010,%
Vityazev2010AIPC,Zotov2010ArtSa,Miller2011,Seitz2012JGRB,Zotov2012JGeo,%
Beutler2020AdSpR,Lopes2022Geosc,Zotov2022MUPB,Yamaguchi2024}.

In this study, we have focused on the 19th century part
of the IERS C01 PM series, which is very important for
analysis of the long-term variations in rotation of the Earth,
in particular, Chandler wobble (CW), and annual wobble (AW).
Although the pole coordinates related to the 19th century have relatively
low accuracy compared to PM data obtained after 1900, they are of great
importance for investigation of long-term variations the Earth rotation.
Suffice it to say that as early as the late 19th century
\citet{Chandler1894_322} separated CW and AW signals, derived
the value of the CW period of 428-430 days, which is close to
the latter determinations, and discovered an $\approx$66-year period
of the CW amplitude variation, which is only about 20\% shorter than
recently obtained from 180-year PM series.
Including the 19th century PM data into modern analyses of the entire
IERS C01 PM series provides good evidence for ~80-year CW periodicity
in amplitude and phase variations, which cannot be reliably estimated
using only observations since 1900.
All this certainly confirms how useful the 19th century PM series is
for studying long-term behavior of the PM.

Unfortunately, the IERS C01 series has a 2-year gap in $Y_p$
pole coordinate in the date range 1858.9 to 1860.9.
Of course, a time series without missing epochs is always
better than a series with a gap.
Missing data can cause problems in statistical data analysis.
In particular, for spectral analysis, most of the standard
methods used require continuous regularly spaced data.
If the latter is the case, the results are free from the risk
of introducing spurious frequencies in the spectrum.
It is also possible to use other efficient analysis methods
for regularly spaced data.
The reconstructed continuous evenly spaced PM series can also
provide new information about the evolution of the Earth's
rotation modes in the gap epochs.

In this paper, for the first time, to our knowledge,
we examine some approaches to address the problem
of filling the gap in the IERS C01 series.
The challenge is to fill in the missing $Y_p$ values in
the series while preserving its structure as much as possible.
For this purpose, two methods of analysis were considered.
The first is the use of a parametric astronomical model,
specially developed for this study and consisted of three components:
bias, CW, and AW (hereafter referred to as BCA model).
The second data-driven model has been constructed
using Singular Spectrum Analysis (SSA) \citep{Golyandina2001}.
These two methods cover the range of models, from parametric
to nonparametric, and allow the treatment of time series that
do not have a single fundamental period and are rather short.

The paper is organized as follows.
Section~\ref{sect:data} describes the data used in this study
and presents the results of its preliminary analysis.
Section~\ref{sect:bca_analysis} is devoted to deriving
a parametric model of the IERS C01 series.
A refined analysis of the IERS C01 PM series using the
SSA technique is performed in Section~\ref{sect:ssa_analysis}.
Details of SSA analysis are described in Appendix A.
Section~\ref{sect:final_results} sums up the results obtained in this work.
Concluding remarks on this study are given in the final Section.

%%%%%%%%%%%%%%%%%%%%%%%%%%%%%%%%%%%%%%%%%%%%%%%%%%%%%%%%%%%%%%%%%%%%%%%%%%%%%%%%%%%%%%%%%%%%%%%%%%%%%%%%%%%%%%

\section{Data description and preliminary analysis}
\label{sect:data}

\begin{figure*}
\centering
\includegraphics[clip,width=\hsize]{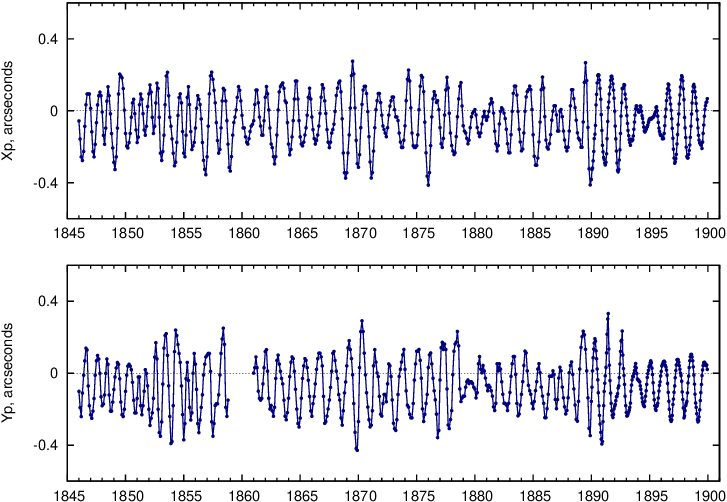}
\caption{The $X_p$ component (top panel) and $Yp$ component (bottom panel) of the IERS C01 series.}
\label{fig:c01_series}
\end{figure*}

The IERS C01 PM series consists of several parts as described
in Table~\ref{tab:c01content}.
The epoch step in this series is 0.1~yr in 1846---1890,
and 0.05~yr starting with 1890.
In this study, the beginning of the IERS C01 series
in the date range 1846 to 1899 was used.
Corresponding PM data are shown in Fig.~\ref{fig:c01_series}

\begin{table}
\centering
\caption{Content of the IERS C01 polar motion series.}
\label{tab:c01content}
\begin{tabular}{ll}
Date interval & Data source \\
\hline
1846 -- 1889 & \citet{Rykhlova1970SoobGAISH} \\
1890 -- 1899 & \citet{Fedorov1972} \\
1900 -- 1961 & \citet{Vondrak1995,Vondrak2010} \\
1962 -- now  & \citet{Bizouard2019}
\end{tabular}
\end{table}

As can be seen from Table~\ref{tab:c01content},
the part of the IERS C01 PM series related to 19th century is based
on two papers by \citet{Rykhlova1970SoobGAISH} and \citet{Fedorov1972}.
While the work of \citet{Fedorov1972} is widely available and well
documented, the original work \citet{Rykhlova1970SoobGAISH} is less
known and available and it makes sense to briefly describe it here.
This work was based on the analysis of latitude determinations that were
made in the framework of several programs for determination of absolute
declinations of stars at three observatories: Greenwich, Pulkovo, and Washington.
The Greenwich latitude variations were taken from
\citet{Chandler1894_320} (years 1836.0--1851.0) and
\citet{Thackeray1893} (years 1851.0--1891.5).
The Pulkovo latitude variations were taken from
\citet{Orlov1961} (years 1842.0--1849.5, 1863.6--1879.0, and 1882.0--1892.0).
The Washington latitude variations were derived in
\citet{Rykhlova1970TrGAISH} (years 1845.1--1858.8 and 1861.0--1888.0).
All three stations were given equal weights.

The pole coordinates were computed using the formula
\begin{equation}
\Delta\varphi = X_p\cos\lambda + Y_p\sin\lambda \,,
\label{eq:phi_pole}
\end{equation}
where $\Delta\varphi$ is the change of the latitude caused by the PM,
and $\lambda$ is the longitude of the observatory.
Before computing the pole coordinates, the mean latitude
was subtracted from all three latitude series.
The same applies to the PM series computed by \citet{Fedorov1972}.
The mean latitude $\bar{\varphi}$ was computed using the formula
applied to latitude time series sampled at 0.1 year:
\begin{equation}
\bar{\varphi}_i = \left(\sum_{i=0}^{9} \varphi_{i-8}+\varphi_{i-7}+\varphi_{i-2}+\varphi_{i-1}\right)/40 \,.
\end{equation}

As a consequence, the trend component is practically absent
in the IERS C01 series before 1900.0.
This can be clearly seen in Fig.~\ref{fig:c01_series} and
is confirmed by further SSA analysis
(see Figs.~\ref{fig:ssa_6components} and~\ref{fig:ssa_9components}).
However, such inconsistency of the two parts of the IERS C01 series
before and after 1900.0 with respect to the secular (trend) component
does not affect the results of the study of CW and AW variations
if an appropriate analysis technique is used.
In our case (see the following sections), we use only the data before 1900.0.
When using the entire IERS C01 series, the trend component can be removed,
for example, using the SSA method or any high-pass filtering.

For most epochs of the Rykhlova's PM series, data from all
three observatories were used.
In the period when Pulkovo was not observing, the pole coordinates were
computed using data from Greenwich and Washington.
In the Rykhlova's series and, consequently, the IERS C01 PM series
there are no $Y_p$ data from 1858.9 to 1860.9 inclusive.
A total of 21 epochs are missing.
The middle epoch of the gap is 1859.9.
The gap is explained by absence of observations at Washington observatory
during this period.
The problem is that the Greenwich and Pulkovo observatories are located
at close longitudes, $0^{\circ}$ and $30^{\circ}$, respectively, which
is close to direction of $X_p$.
Therefore, it is impossible to accurately estimate the both pole
coordinates using only observations from these two stations.
Consequently, only $X_p$ values were computed for these epochs.
The Washington observatory is located at longitude $-77^{\circ}$,
which is close to the $Y_p$ direction, and its inclusion in the processing
provides a good longitude coverage to reliably determine both pole coordinates.

\begin{figure}
\centering
\includegraphics[clip,width=0.5\hsize]{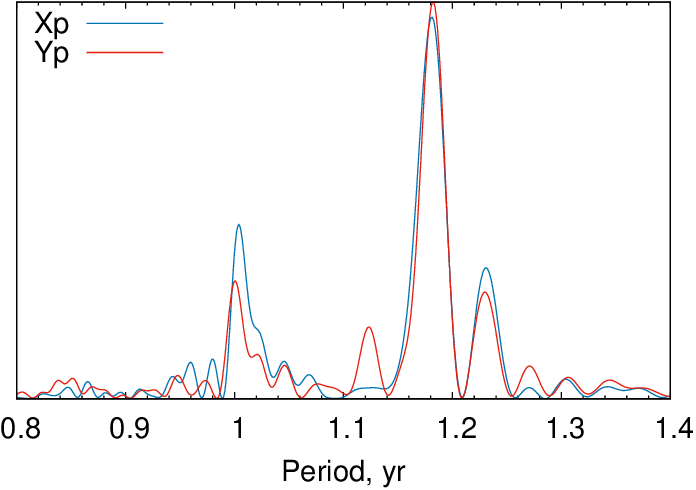}
\caption{Generalized Lomb-Scargle periodogram of the IERS C01 PM series
  for the interval 1846--1899, arbitrary units.}
\label{fig:c01_spectra}
\end{figure}

Although the polar motion variations described by the IERS C01
series for 1846--1899 are assumed to consist mainly of AW and CW
components, possibly with
bias\footnote{\url{ftp://hpiers.obspm.fr/iers/eop/eopc01/EOPC01.GUIDE}},
spectral analysis has shown that in fact the signal has a more
complex structure as shown in Fig.~\ref{fig:c01_spectra}.
The generalized Lomb-Scargle periodograms
for the C01 $X_p$ and $Y_p$ series were computed according to
\citet{Zechmeister2009} after removing the bias clearly visible
in Fig.~\ref{fig:c01_series}.

%%%%%%%%%%%%%%%%%%%%%%%%%%%%%%%%%%%%%%%%%%%%%%%%%%%%%%%%%%%%%%%%%%%%%%%%%%%%%%%%%%%%%%%%%%%%%%%%%%%%%%%%%%%%%%

\section{Parametric model}
\label{sect:bca_analysis}

It is not possible to use a straightforward model consisting
of two components, CW and AW, because their amplitudes do not
remain constant over time in the date range around the gap
\citep{Kolaczek1999srst,Yatskiv2000ASPC,Guo2005JGeod,Miller2011,Lopes2022Geosc,Zotov2022MUPB}.
However, these two independent analyses of the IERS C01 series
showed that the CW amplitude grew nearly linearly in 1850--1870.
\citet{Miller2011} also showed that the AW amplitude is nearly
constant in this period with a tiny increase over time.
Therefore, it is possible to construct an approximation model
for $X_p$ and $Y_p$ for the period around 1850--1870,
consisting of a bias and two oscillations with CW and AW
periods and linear amplitude trend:
\begin{equation}
\begin{array}{rcl}
Y_p &=& a_0 \\
  &+& (1+a_1^c t)
    \left(a_c^c\cos\fracd{2\pi(t-t_0)}{P_c}+a_s^c\sin\fracd{2\pi(t-t_0)}{P_c} \right)\\[2ex]
  &+& (1+a_1^a t)
    \left(a_c^a\cos\fracd{2\pi(t-t_0)}{P_a}+a_s^a\sin\fracd{2\pi(t-t_0)}{P_a} \right) \,,
\end{array}
\label{eq:approximation_cwaw}
\end{equation}
or, equivalently,
\begin{equation}
\begin{array}{rcl}
Y_p &=& a_0 \\[1ex]
  &+& a_{c0}^c\cos\fracd{2\pi(t-t_0)}{P_c}+a_{s0}^c\sin\fracd{2\pi(t-t_0)}{P_c} \\[2ex]
  &+& a_{c1}^ct\cos\fracd{2\pi(t-t_0)}{P_c}+a_{s1}^ct\sin\fracd{2\pi(t-t_0)}{P_c} \\[2ex]
  &+& a_{c0}^a\cos\fracd{2\pi(t-t_0)}{P_a}+a_{s0}^a\sin\fracd{2\pi(t-t_0)}{P_a} \\[2ex]
  &+& a_{c1}^at\cos\fracd{2\pi(t-t_0)}{P_a}+a_{s1}^at\sin\fracd{2\pi(t-t_0)}{P_a} \,,
\end{array}
\label{eq:approximation_cwaw_2}
\end{equation}
where $t$ is epoch in years, $t_0$=1859.9 (the middle epoch of the gap),
$P_c$=1.19~yr is the period of the CW, and $P_a$=1~yr is the period of the AW.

\begin{figure*}
\centering
\includegraphics[clip,width=\textwidth]{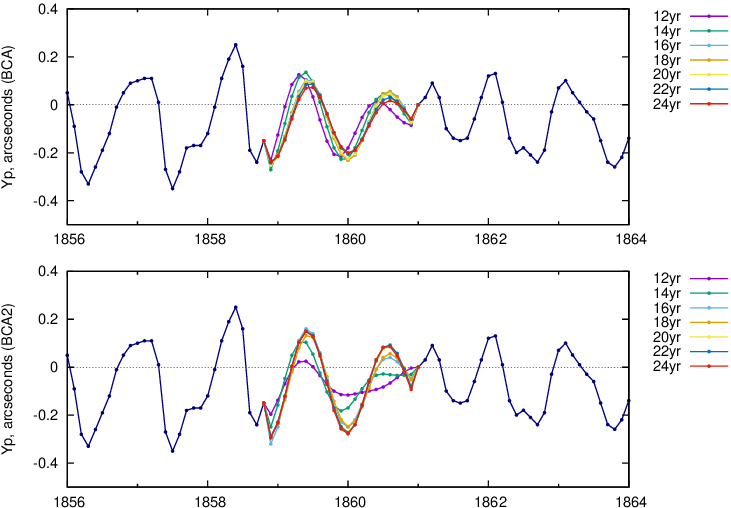}
\caption{IERS C01 $Y_p$ series with filled missed data
  between 1858.9 and 1860.9 using the BCA model (upper panel)
  and BCA2 (bottom panel).
  Plot titles show date intervals used to fit the $Y_p$ data
  to the model.}
\label{fig:filling_gap_bca}
\end{figure*}

The nine parameters of this model were fitted using the least squares method (LS)
over seven intervals of length of 12 to 24 years with a step of 2 years centered
at the middle epoch of the gap in $Y_p$ series, 1859.9, in total seven variants.
Then the filling $Y_p$ values were computed using this model for each variant.
The results of these computations are shown in Fig.~\ref{fig:filling_gap_bca} (upper panel).

The BCA model represented by Eq.~\ref{eq:approximation_cwaw_2}
corresponds to the PM model with a single dominant CW frequency,
which is generally accepted at present.
However, the PM spectrum presented in Fig.~\ref{fig:c01_spectra}
reveals two peaks in the CW frequency band with periods of
1.181~yr and 1.231~yr.
There may be various reasons for this bifurcation of the PM spectrum,
such as an inconsistency in the data on which the series is based,
or CW phase variations.
In any case, it would be interesting to check how this effect may
affect the result of the computation of the modeled $Y_p$ in the gap.

For this purpose, the BCA2 model was constructed as follows:
\begin{equation}
\begin{array}{rcl}
Y_p &=& a_0 \\[1ex]
  &+& a_{c0}^{c1}\cos\fracd{2\pi(t-t_0)}{P_{c1}}+a_{s0}^{c1}\sin\fracd{2\pi(t-t_0)}{P_{c1}} \\[2ex]
  &+& a_{c1}^{c1}t\cos\fracd{2\pi(t-t_0)}{P_{c1}}+a_{s1}^{c1}t\sin\fracd{2\pi(t-t_0)}{P_{c1}} \\[2ex]
  &+& a_{c0}^{c2}\cos\fracd{2\pi(t-t_0)}{P_{c2}}+a_{s0}^{c2}\sin\fracd{2\pi(t-t_0)}{P_{c2}} \\[2ex]
  &+& a_{c1}^{c2}t\cos\fracd{2\pi(t-t_0)}{P_{c2}}+a_{s1}^{c2}t\sin\fracd{2\pi(t-t_0)}{P_{c2}} \\[2ex]
  &+& a_{c0}^a\cos\fracd{2\pi(t-t_0)}{P_a}+a_{s0}^a\sin\fracd{2\pi(t-t_0)}{P_a} \\[2ex]
  &+& a_{c1}^at\cos\fracd{2\pi(t-t_0)}{P_a}+a_{s1}^at\sin\fracd{2\pi(t-t_0)}{P_a} \,,
\end{array}
\label{eq:approximation_cwaw_3}
\end{equation}
where $t$ is epoch in years, $t_0$=1859.9 (the middle epoch of the gap),
$P_{c1}$=1.181~yr is the period of the first CW component,
$P_{c2}$=1.231~yr is the period of the second CW component,
and $P_a$=1~yr is the period of the AW.
In this case, a longer interval of dates is needed to separate the two
CW components.
The 13 parameters of this model were fitted using the least squares method (LS)
over seven intervals of length of 12 to 24 years with a step of 2 years
centered at the middle epoch of the gap in $Y_p$ series, 1859.9,.
The filling $Y_p$ values for the BCA2 model are shown
in Fig.~\ref{fig:filling_gap_bca} (bottom panel).
They are close to the results obtained with the BCA model.

To evaluate the real accuracy of the PM series approximation by the parametric models,
we computed the differences between the modeled and observed PM values
for each of seven variants for the epochs before and after the gap.
Then we computed the root mean square error (RMSE) and mean absolute error (MAE)
for these differences.
The same computations were repeated for the $X_p$ series, for which we used the original
IERS C01 series from which the epochs 1858.9--1860.9 were cut to make it similar
to the $Y_p$ series.
The results of this error analysis are shown in Table~\ref{tab:bca_errors}.
Two main observations can be made from this table.
First, shorter intervals provide better approximation, as expected.
Secondly, the approximation error for the $X_p$ series is better than for the $Y_p$
series, which corresponds to a higher precision of the $X_p$ values with respect
to the $Y_p$ values given in the IERS C01 series in the 19th century.
Although the approximation error for the BCA2 model is better than that for the BCA model,
the latter may be more significant from a scientific point of view because a duality
of the CW period is doubtful.
For this reason, the average of two first BCA variants and two first BCA2 variants,
i.e. those obtained for 12-year and 14-year variants, hereafter referred to as BCA model, will be used for further comparisons.

\begin{table}
\centering
\caption{Approximation error for the parametric models BCA and BCA2, arcseconds.}
\label{tab:bca_errors}
\begin{tabular}{ccccc}
\hline
Interval of dates & \multicolumn{2}{c}{$X_p$} & \multicolumn{2}{c}{$Y_p$} \\
& MAE & RMSE & MAE & RMSE \\
\hline
\multicolumn{5}{c}{BCA model} \\
1853.89--1865.91 & 0.068 & 0.081 & 0.067 & 0.090 \\
1852.89--1866.91 & 0.066 & 0.081 & 0.071 & 0.095 \\
1851.89--1867.91 & 0.079 & 0.099 & 0.084 & 0.110 \\
1850.89--1868.91 & 0.077 & 0.096 & 0.083 & 0.108 \\
1849.89--1869.91 & 0.078 & 0.097 & 0.084 & 0.111 \\
1848.89--1870.91 & 0.079 & 0.098 & 0.087 & 0.112 \\
1847.89--1871.91 & 0.079 & 0.098 & 0.085 & 0.110 \\
\multicolumn{5}{c}{BCA2 model} \\
1853.89--1865.91 & 0.062 & 0.076 & 0.060 & 0.084 \\
1852.89--1866.91 & 0.064 & 0.078 & 0.060 & 0.085 \\
1851.89--1867.91 & 0.073 & 0.091 & 0.070 & 0.095 \\
1850.89--1868.91 & 0.073 & 0.092 & 0.076 & 0.102 \\
1849.89--1869.91 & 0.071 & 0.089 & 0.077 & 0.102 \\
1848.89--1870.91 & 0.071 & 0.088 & 0.075 & 0.099 \\
1847.89--1871.91 & 0.073 & 0.092 & 0.076 & 0.100 \\
\hline
\end{tabular}
\end{table}

Generally speaking, other advanced models can be constructed
in a similar way with more complicated behavior of trend,
and CW and AW amplitudes (and maybe phases), but such
an extension does not look reasonable in our case because
there is no data enough to fit a many-parameter model.
Therefore, analysis based on the BCA model cannot provide
a fully satisfactory solution because it relies on a model
that only approximately describes the actual pole motion.
However, it may be useful for verification of other models.
The more reliable solution to the problem to which this study
is aimed will be obtained in the next Section using
the data-driven SSA technique.

%%%%%%%%%%%%%%%%%%%%%%%%%%%%%%%%%%%%%%%%%%%%%%%%%%%%%%%%%%%%%%%%%%%%%%%%%%%%%%%%%%%%%%%%%%%%%%%%%%%%%%%%%%%%%%

\section{SSA analysis}
\label{sect:ssa_analysis}

The SSA method is one of the most powerful tools for analyzing
time series with a complex structure including the PM series
\citep{Vautard1989,Elsner1996,Golyandina2001,Ghil2002,Vityazev2010AIPC,Miller2011,Golyandina2020}.
It is widely used for analyzing geodetic, geophysical, and
astronomical time series, in particular, for filling gaps in time series
\citep{Kondrashov2006,Golyandina.Osipov2007,Shen2015,Golyandina2019,Yi2021,Ji2023}.
SSA is an agile method because it can propose a (possibly local) approximation by a model
in the form of a finite sum of products of polynomials, exponentials and harmonics.
This model is called linear because it consists of signals governed by linear recurrence relations.
However, this is a rather large class of signals, and a signal may only approximately satisfy the model.
This means that within a two-year time frame, this property of SSA overcomes its limitations.
It should be noted that the set of frequencies does not change during the time
interval considered, but rather their amplitudes.
By varying the window length, SSA can account for amplitude changes even
in the case of some non-linearity.

For further presentation of our method, we denote IERS C01
PM coordinates $X_p$ and $Y_p$ as $\tX$ and $\tY$, respectively.
As in the case of the parametric model considered in Section~\ref{sect:bca_analysis},
we will consider a model in the form of signal plus noise, assuming that the signal
consists of three main components: trend, AW, and CW, with possible addition of other
smaller components.
Unlike parametric model specification, in the SSA method, the model does not need
to be specified in advance, including setting the component periods.
The signal components are defined adaptively by the SSA method and, as will be shown
further on, are indeed composed of the listed components.
We will process the PM time series with 10 measurements per year.
Thus, the AW component corresponds to a period of 10 points,
and the CW component corresponds to a period of about 11.9 points.
Unlike the parametric BCA model described above, the SSA analysis
does not imply a priori restrictions on the trend, AW and CW variations.
The general approach to gap filling with SSA is described in Appendix~A.

In the real form, the SSA method works well with signals as a sum
of possibly modulated harmonics \citep{Golyandina2001}.
Such a modification of SSA applied to a single time series $\tY$
will be called hereafter 1D-SSA.

It was shown in \citet{Golyandina.etal2015} that if several
signals have the same periods (with the same modulation),
Multivariate SSA (MSSA), a generalization of SSA for
time series system analysis, extracts the signal more accurately.
Such a method allows us to process $\tX$ and $\tY$ series
together.

Taking into account the physical nature of the pole coordinates,
this looks reasonable to consider a representation of the PM
series as a complex-valued time series and applied Complex SSA (CSSA).
If the input signal consists of two harmonics with a phase
shift of $\pi/2$, then CSSA extracts the signal even
more accurately \citet{Golyandina.etal2015}.

To ensure more reliable results, we considered the three methods
described above for reconstruction of PM signal:
basic 1D-SSA for the series $\tY$, MSSA for the series
($\tX, \tY$), and CSSA for the series $\tX+\iu\tY$.
Note that MSSA and CSSA are applied to both time series, $\tX$ and $\tY$,
together, so they take into account the relationships between the series.
The methods MSSA and CSSA result in decompositions of both series,
but we will use only the part of the decomposition of the series Y.

An important option for a practical SSA realization is a choice
of the window length.
Recommendations for optimal window selection include
proportionality to the periods of the harmonics
we want to extract \citep{Golyandina2001}.
Therefore, for the window length $L$ we will consider three
values, 60, 119 and 238 points, since the AW and CW periods,
1.0 and 1.19~yr, respectively, are included near integer
number of times in windows of this length.

The parameter that we want to optimize using artificial gaps
is $r$, the number of the SVD components assigned to the signal.
We will call $r$ the model order.
To find the optimal value of the model order, we vary it
from 1 to 17 and calculate the error of gap filling
according to Algorithm~1 (Appendix 1).
We again used RMSE and MAE as a measure of the difference between
the imputed values and the input time series values.

As a set of artificial gaps, we considered $R=120$ gap intervals
of length $M=21$, the same as the actual gap.
The first interval starts at 1846.1, the remaining 119 intervals
are obtained by successively shifting the initial interval four
points forward (intervals with actual gaps were excluded).

The mean MAE and RMSE errors for these models are presented
in Table~\ref{tab:ssa_errors}.
These test results have shown that the most precise filling
of artificial gaps were achieved using CSSA with
($L$=60, $r$=4, 5), ($L$=119, $r$=6--9), ($L$=238, $r$=7).
These seven variants provide practically the same error of the approximation
of the $\tY$ series.
Similar results were obtained for RMSE, for which the optimal parameters
$r$ and $L$ turned out to be the same as for the MAE estimate.
Thus, it has been found that the obtained conclusions do not depend
on the choice of the method for estimating the modeling error.

\begin{table*}
\centering
\caption{MAE/RMSE filling errors on artificial gaps
  for SSA model order $r = 1 \ldots 17$, arcseconds.}
\label{tab:ssa_errors}
\begin{tabular}{rccccccccc}
\hline
$r$ & \multicolumn{3}{c}{$L$=60} & \multicolumn{3}{c}{$L$=119} & \multicolumn{3}{c}{$L$=238} \\
   & 1D-SSA    & MSSA      & CSSA               & 1D-SSA    & MSSA      & CSSA               & 1D-SSA    & MSSA      & CSSA \\
\hline
 1 & .125/.150 & .117/.145 &         .104/.129  & .121/.143 & .114/.142 &         .104/.130  & .121/.143 & .114/.142 &         .106/.135  \\
 2 & .112/.138 & .114/.142 &         .084/.109  & .104/.129 & .109/.137 &         .084/.110  & .107/.134 & .108/.137 &         .087/.115  \\
 3 & .093/.118 & .101/.124 &         .070/.092  & .089/.116 & .094/.118 &         .076/.099  & .091/.120 & .091/.118 &         .080/.105  \\
 4 & .081/.107 & .102/.124 & \textbf{.066/.085} & .082/.109 & .090/.113 &         .072/.094  & .085/.113 & .086/.112 &         .076/.099  \\
 5 & .080/.107 & .110/.134 & \textbf{.064/.083} & .085/.110 & .092/.114 &         .070/.090  & .086/.112 & .086/.111 &         .072/.094  \\
 6 & .078/.105 & .116/.140 &         .067/.087  & .083/.108 & .095/.117 & \textbf{.065/.084} & .083/.106 & .086/.109 &         .071/.091  \\
 7 & .081/.109 & .124/.150 &         .067/.088  & .085/.111 & .101/.123 & \textbf{.064/.084} & .084/.106 & .087/.109 & \textbf{.066/.085} \\
 8 & .078/.103 & .131/.158 &         .071/.091  & .086/.112 & .106/.129 & \textbf{.064/.084} & .083/.106 & .090/.112 &         .073/.093  \\
 9 & .083/.107 & .139/.168 &         .073/.095  & .090/.117 & .113/.137 & \textbf{.065/.085} & .084/.109 & .096/.118 &         .075/.096  \\
10 & .086/.111 & .144/.174 &         .077/.099  & .088/.113 & .116/.139 &         .068/.090  & .085/.109 & .100/.122 &         .076/.097  \\
11 & .094/.121 & .148/.179 &         .079/.102  & .088/.111 & .118/.143 &         .071/.092  & .085/.110 & .104/.126 &         .074/.094  \\
12 & .092/.119 & .150/.182 &         .083/.107  & .080/.102 & .123/.149 &         .072/.095  & .083/.107 & .105/.127 &         .074/.095  \\
13 & .095/.124 & .152/.184 &         .085/.110  & .081/.106 & .129/.156 &         .073/.097  & .084/.108 & .108/.130 &         .076/.098  \\
14 & .097/.126 & .153/.186 &         .087/.112  & .085/.111 & .133/.160 &         .073/.099  & .085/.108 & .110/.133 &         .076/.099  \\
15 & .103/.133 & .155/.188 &         .087/.110  & .090/.118 & .137/.166 &         .071/.097  & .090/.114 & .114/.137 &         .070/.091  \\
16 & .105/.136 & .156/.189 &         .085/.107  & .093/.121 & .140/.169 &         .075/.102  & .086/.110 & .116/.139 &         .070/.091  \\
17 & .110/.143 & .157/.191 &         .094/.116  & .092/.122 & .143/.173 &         .081/.108  & .088/.115 & .119/.142 &         .070/.091  \\
\hline
\end{tabular}
\end{table*}

\begin{figure*}
\centering
\includegraphics[clip,width=\hsize]{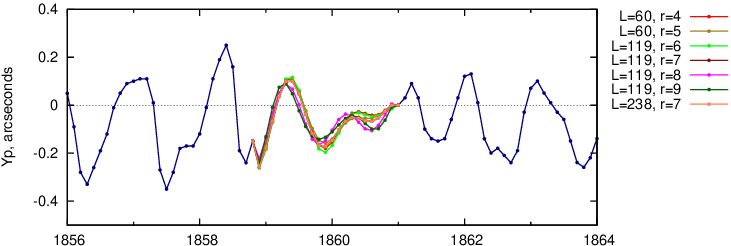}
\caption{IERS C01 $Y_p$ series with filled missed data
  between 1858.9 and 1860.9 computed using the CSSA model with
  the seven models providing the minimal approximation error.}
\label{fig:filling_gap_ssa}
\end{figure*}

Comparison of the data presented in Table~\ref{tab:bca_errors} and
Table~\ref{tab:ssa_errors} shows that the accuracy of the $Y_p$ series
approximation using CSSA method is a bit better than that for the BCA method.
Remarkably, that the best RMSE in filling $\tY$ values achieved by both methods
(0.08$''$--0.09$''$) are very close to the $Y_p$ uncertainty given
in the IERS C01 series (0.09$''$), which is assumed to be the standard
deviation, equivalent to RMSE), confirming both the reality
of the IERS C01 errors and the correctness of our error estimate.
The gap-filling plots with the best parameters described above
are shown in Fig.~\ref{fig:filling_gap_ssa}.

\begin{figure*}
\centering
\includegraphics[clip,width=\textwidth]{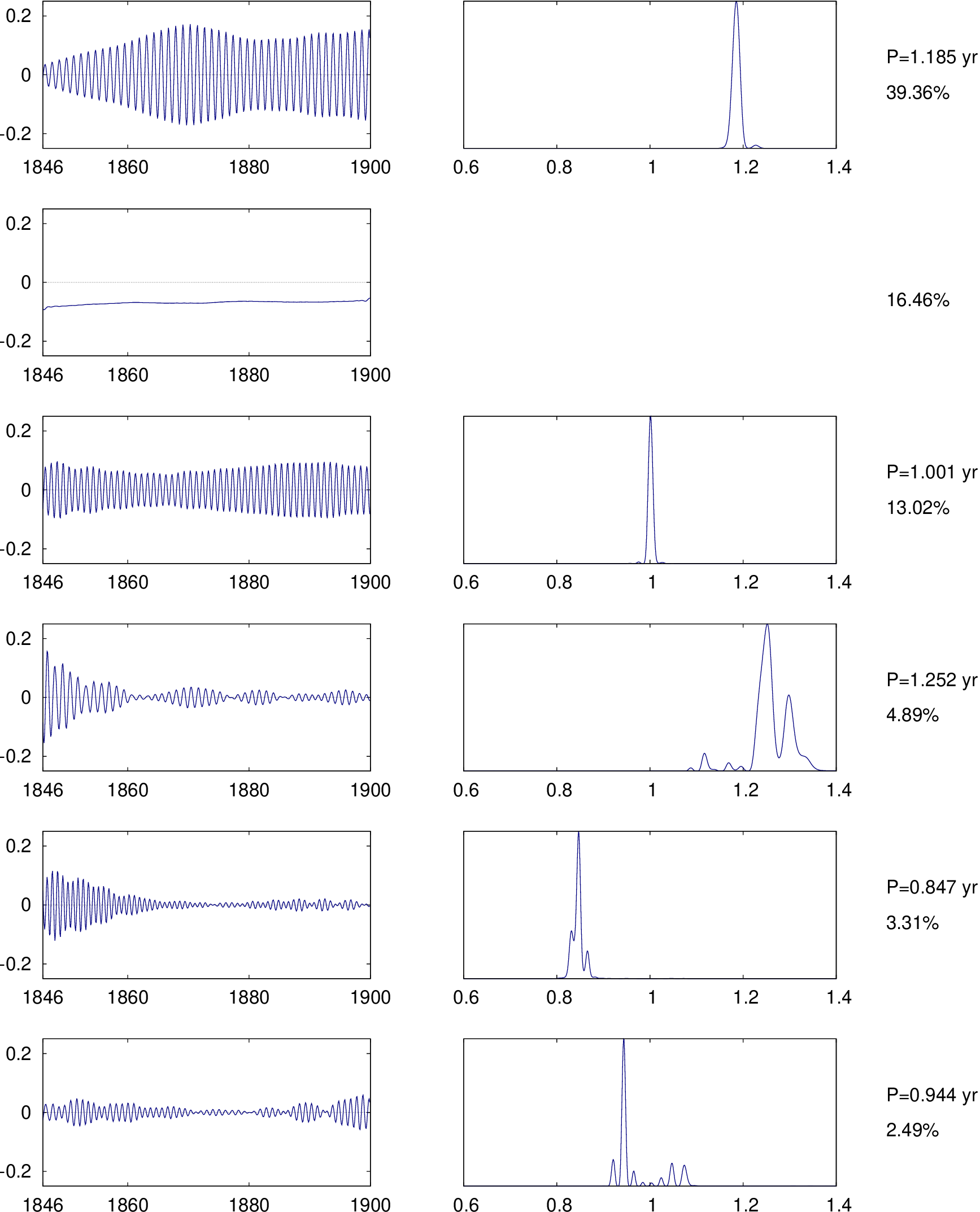}
\caption{The $\tY$ components of CSSA solution with $r$=6 (in arcseconds)
  and their Lomb-Scargle periodograms (except the trend component, arbitrary units).
  The period of the most powerful oscillation and the contribution
  of the component to the total signal are shown on the right.}
\label{fig:ssa_6components}
\end{figure*}

\begin{figure*}
\centering
\includegraphics[clip,width=\textwidth]{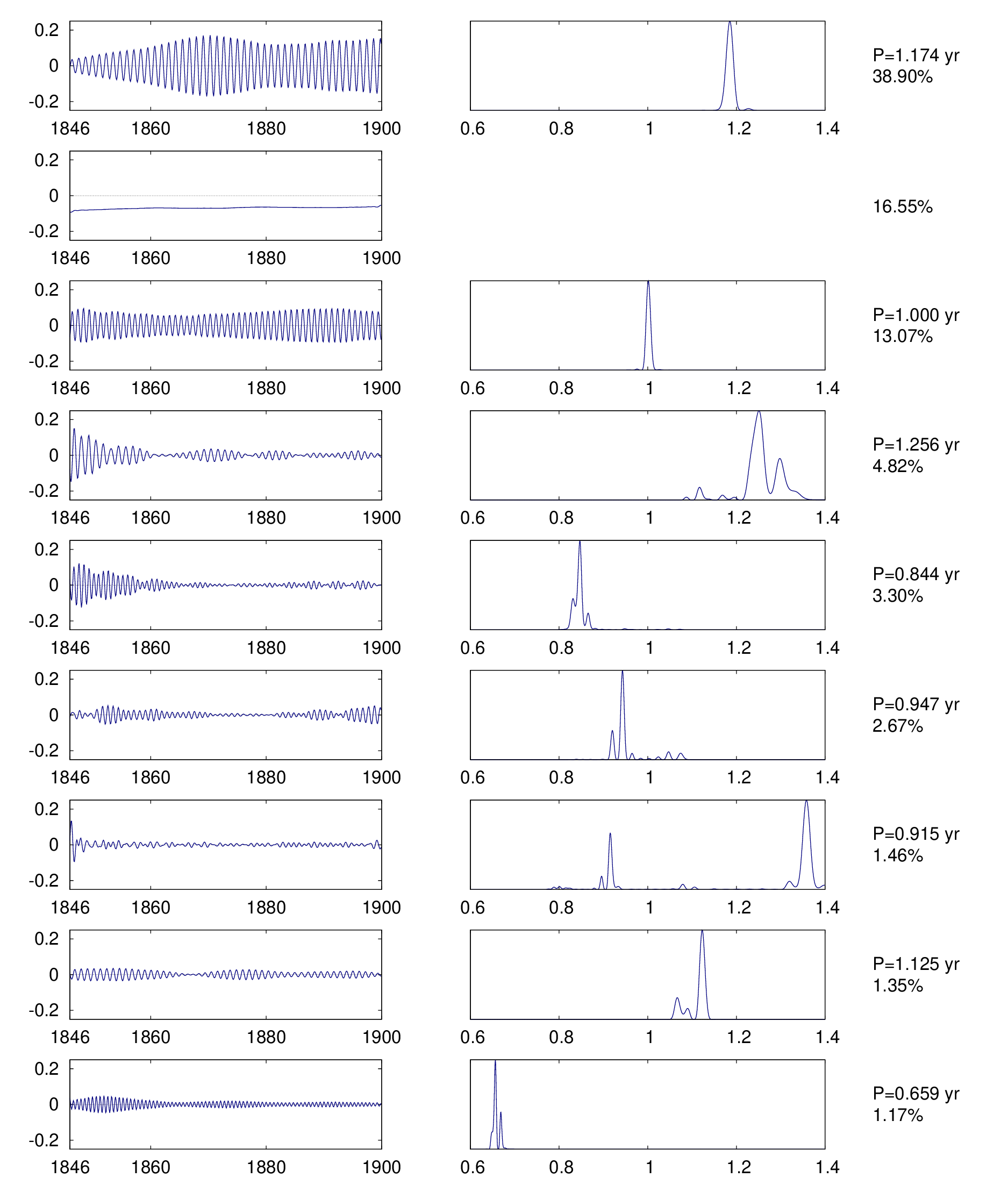}
\caption{The $\tY$ components of CSSA solution with $r$=9 (in arcseconds)
  and their Lomb-Scargle periodograms (except the trend component, arbitrary units).
  The period of the most powerful oscillation and the contribution
  of the component to the total signal are shown on the right.}
\label{fig:ssa_9components}
\end{figure*}

Consider the CSSA decomposition components of the time series
$\tY$ after filling the actual gap.
Recall that the imaginary parts of the complex components
are related to the series $\tY$.
The latter for the CSSA 6-component solution with $L=119$
are shown in Fig.~\ref{fig:ssa_6components}, and
Fig.~\ref{fig:ssa_9components} shows the $\tY$ components
for the 9-component CSSA solution, as examples.

It can be seen that the leading six components of the 9-component
solution have practically the same periods as those for the
6-component solution. Some components show two spectral peaks,
which can be caused by mixing of two periodicities
or by phase instability of the component.
In all solutions, the first component corresponds to the main
CW oscillation, and the third component corresponds to the
main AW oscillation.
In Figs.~\ref{fig:ssa_6components} and \ref{fig:ssa_9components},
two minor components with periods of 0.847~yr and 0.944~yr can also be seen.
The corresponding peaks in Fig.~\ref{fig:c01_spectra} are small, at the noise level.
We consider this fact as confirmation of the better sensitivity of the SSA method
to small components in the analyzed signal, which allows for their better detection.

Since the imputed values for the seven CSSA models described above
provide approximately equal errors on the artificial gaps, the average
of these values can be recommended as the final method of filling in
the actual gap in the IERS C01 series.
This filling method using the average of seven gap fillings has a slightly
smaller MAE (0.0628 vs. 0.0638--0.0662) on the considered set of artificial gaps,
compared to each of these seven gap fillings.

%%%%%%%%%%%%%%%%%%%%%%%%%%%%%%%%%%%%%%%%%%%%%%%%%%%%%%%%%%%%%%%%%%%%%%%%%%%%%%%%%%%%%%%%%%%%%%%%%%%%%%%%%%%%%%

\section{Final results}
\label{sect:final_results}

\begin{figure*}
\centering
\includegraphics[clip,width=\hsize]{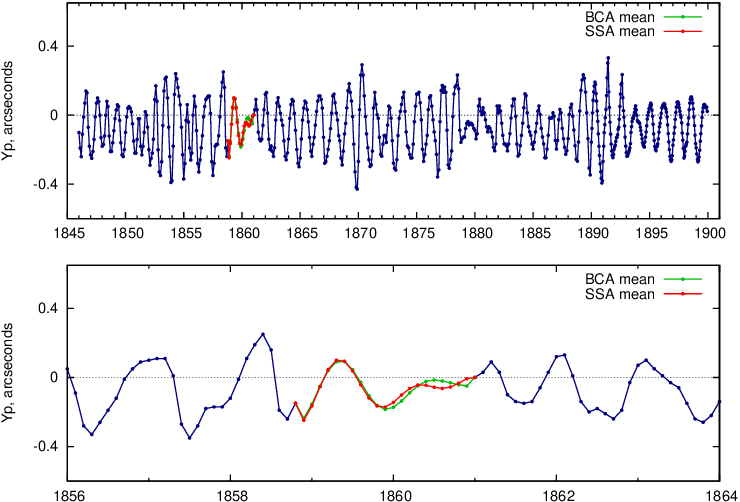}
\caption{IERS C01 $Y_p$ series with filled missed data
  between 1858.9 and 1860.9 using the average values
  for both BCA and SSA models.}
\label{fig:filling_gap}
\end{figure*}

Average results (filling $Y_p$ values) computed using the BCA
and SSA methods are presented in Table~\ref{tab:y_gap_results}
with the number of decimal places corresponding to the IERS C01 series.
The IERS C01 series with added filling values is shown in
Fig.~\ref{fig:filling_gap}.

\begin{table}
\begin{center}
\caption{Filled $Y_p$ values, arcseconds.}
\label{tab:y_gap_results}
\begin{tabular}{crrr}
\hline
Epoch & \multicolumn{2}{c}{Model} & \multicolumn{1}{c}{Mean} \\
& \multicolumn{1}{c}{BCA} & \multicolumn{1}{c}{SSA} & \\
\hline
1858.9 & $-0.236505$ & $-0.247675$ & $-0.242090$ \\
1859.0 & $-0.153508$ & $-0.164798$ & $-0.159153$ \\
1859.1 & $-0.050599$ & $-0.053550$ & $-0.052074$ \\
1859.2 & $ 0.039868$ & $ 0.045943$ & $ 0.042905$ \\
1859.3 & $ 0.091581$ & $ 0.100224$ & $ 0.095903$ \\
1859.4 & $ 0.092240$ & $ 0.095042$ & $ 0.093641$ \\
1859.5 & $ 0.046379$ & $ 0.038934$ & $ 0.042656$ \\
1859.6 & $-0.027558$ & $-0.042570$ & $-0.035064$ \\
1859.7 & $-0.104471$ & $-0.118523$ & $-0.111497$ \\
1859.8 & $-0.161363$ & $-0.164565$ & $-0.162964$ \\
1859.9 & $-0.184487$ & $-0.170918$ & $-0.177702$ \\
1860.0 & $-0.172543$ & $-0.143915$ & $-0.158229$ \\
1860.1 & $-0.135271$ & $-0.101195$ & $-0.118233$ \\
1860.2 & $-0.088418$ & $-0.063027$ & $-0.075722$ \\
1860.3 & $-0.047375$ & $-0.043324$ & $-0.045350$ \\
1860.4 & $-0.021975$ & $-0.044098$ & $-0.033036$ \\
1860.5 & $-0.014260$ & $-0.055752$ & $-0.035006$ \\
1860.6 & $-0.019681$ & $-0.063258$ & $-0.041469$ \\
1860.7 & $-0.030755$ & $-0.055410$ & $-0.043083$ \\
1860.8 & $-0.041278$ & $-0.032369$ & $-0.036824$ \\
1860.9 & $-0.049166$ & $-0.006942$ & $-0.028054$ \\
\end{tabular}
\end{center}
\end{table}

Both methods show good agreement, especially in the first year
of the gap interval, 1859--1860.
The agreement in the second year of the interval, 1860--1861,
is somewhat worse, but can be considered quite satisfactory,
considering that the $Y_p$ error in the IERS C01 series is 0.09$''$.
Generally speaking, it should be kept in mind that the accuracy
of the reconstruction is limited by the accuracy of the IERS C01
data, especially before the gap in the 1840s--1850s.

As was discussed above, the RMSE of the filling $Y_p$ values
obtained for both BCA and SSA methods (0.08$''$--0.09$''$) are very close
to the $Y_p$ uncertainty in the IERS C01 series (0.09$''$).
Therefore, it looks reasonable to assign the IERS C01 $Y_p$ errors
$0.09''$ to the filling values given in Table~\ref{tab:y_gap_results}.

%%%%%%%%%%%%%%%%%%%%%%%%%%%%%%%%%%%%%%%%%%%%%%%%%%%%%%%%%%%%%%%%%%%%%%%%%%%%%%%%%%%%%%%%%%%%%%%%%%%%%%%%%%%%%%

\section{Conclusions}
\label{sect:conclusions}

In this paper, the first attempt has been made to address the problem
of filling the 2-year gap in the IERS C01 $Y_p$ series in 1858.9--1860.9.
For this purpose, two methods were proposed and investigated.
The first is the LS adjustment of the 9-parameter BCA model, consisting
of a bias and two harmonic terms, CW and AW, with linearly changing
amplitude, and its modification, the 13-parameter BCA2 model, which
includes two CW components.
The second method is CSSA (complex SSA), applied to the IERS C01 PM series
in complex form, which allows us to obtain a fully data-driven solution,
not limited by the parameters of parametric models.
Besides, the discrepancies between the results obtained by the
two methods may be due to the fact that the parametric models
are based on a several times shorter interval than the SSA model.
Thus, parametric models mainly describe the local PM behaviour
on the date interval around the gap and under the assumption of
a linear change in the CW and AW amplitudes, while the SSA model
is much less sensitive to a priori assumptions and is more suitable
for describing the long-term PM features, which is the primary
interest of this paper.

Therefore, we consider the result obtained by SSA as more
reliable because it takes into account all the features
of the observed PM data and all the main components of
its structure, while the parametric model uses only
predefined components, which may not adequately describe
the results of the observations.
However, we present both the results obtained by the BCA and SSA
methods in Table~\ref{tab:y_gap_results} so that the user can
choose the variant that looks most appropriate for the problem
under investigation, or simply use the average of the two.

It should be noted that there are many methods for efficiently
processing unevenly spaced data, in particular, data with gaps.
Many of widely used methods of data analysis, such as LS adjustment,
spectral analysis, SSA, wavelet technique, etc., have modifications
for processing unevenly spaced data.
Many examples of application of such analyses to study PM series can be found
in the literature, e.~g., \citet{Gaposchkin1972,Kolaczek1999srst,%
Gambis2000ASPC,Guo2005JGeod,Zotov2012JGeo,Beutler2020AdSpR,Lopes2022Geosc}.
However, the missing two years of $Y_p$ data may introduce bias
in PM modeling if inappropriate analysis methods are used.
Our work aims to provide interested users with the option
to work with evenly distributed data and to use appropriate
mathematical methods, if they want.
In most cases, one cannot expect a significant difference in
the results  obtained with original and filled series,
but working with evenly spaced data is more convenient and
allows for detailed epoch-by-epoch analysis.
Including the proposed filling values into the IERS C01 series
(indeed, with corresponding note) will provide full continuous,
evenly spaced PM series over the entire date range starting from 1846.
This will also allow to avoid (even small) mistakes in using this series
for scientific analysis if the gap in the series in its current,
formally evenly spaced form is not properly taken into account.
Perhaps for this reason, e.~g., \citet{Schuh2001JGeod} and
\citet{Vondrak1988IAUS} started their analysis from 1861.0.

In this study, mathematical methods were mainly used to complete
the existing IERS C01 PM series by filling in the missing $Y_p$ values
for 21 epochs in the period 1858.9 to 1860.9.
Unfortunately, obtained result cannot be verified using the observations
employed by \citet{Rykhlova1970SoobGAISH}.
During the gap period, observations were conducted only by the Greenwich Observatory.
However, Greenwich is located at zero longitude, and therefore the Greenwich
latitude variations are simply equal to $X_p$ (see Eq.~(\ref{eq:phi_pole})),
and are thus insensitive to $Y_p$.

Generally speaking, attempts to improve the quality of the IERS C01 series
in the 19th century, including its extension to the past, should definitely
be considered an actual task.
In particular, mining and reprocessing of historical observations
looks very interesting and important, but this huge task goes far beyond
the scope of this paper and will obviously require cooperation between
observatories that could preserve in their archives declination
and latitude observations carried out in the 19th century.
It should also be mentioned that there have been several studies, such as
\citet{LiXuLin1974,Sekiguchi1975}, attempting to extend the PM series
to earlier dates relative to the beginning of the IERS C01 series,
but these data do not yet seem reliable enough.

On the other hand, extending the IERS C01 PM series at least 20--25 years
into the past, if this is possible at all with acceptable accuracy,
is of special interest.
The Chandler component of PM shows a minimum amplitude and a phase jump
at the epochs around the 1840s, i.e. at the beginning of the IERS C01 series
\citep{Malkin2010,Miller2011}.
Therefore, searching for new observational data and refining the pole coordinates for the period 1820--1860
remains a highly desirable goal for improving our knowledge of the features of the CW variations.

%%%%%%%%%%%%%%%%%%%%%%%%%%%%%%%%%%%%%%%%%%%%%%%%%%%%%%%%%%%%%%%%%%%%%%%%%%%%%%%%%%%%%%%%%%

\section*{Appendix A Filling the gaps with SSA}

Suppose there is a method that knows how to remove noise from a noisy signal.
Then we can propose a simple iterative gap-filling method.
First, the places with missing values are filled with some initial values,
e.~g., the mean of the series, then the signal extraction is performed,
the obtained signal estimates are put at the gaps, the signal extraction
method is applied again, and so the procedure is repeated until convergence.

Typically, the signal estimation method has some parameters.
Then the approach is as follows.
An artificial gap is placed in the part of the series $\tY$
that does not contain gaps and the parameters are chosen so
that the gap filling error is minimized.
For the stability of the result, the error of gap filling
is calculated for different locations of gaps and the errors
are averaged.
It is natural to choose the type of artificial gaps the same
as the actual one. For example, if the actual gap consists
of consecutive missing values and has length $M$, then
artificial gaps should be the same.

\bigskip
\noindent\textbf{Algorithm 1}
(Finding the filling error on artificial gaps):\\[1ex]
\noindent \textit{Input parameters}: time series $\tY$ with a gap,
  length $M$ of artificial gap interval, set of intervals with
  artificial gaps of size $R$, measure of filling errors.
\begin{enumerate}
\item Take an interval of length $M$ from the set of artificial
  gaps and replace the values of the series $\tY$ in this
  interval by missing values.
\item Fill the artificial gap from step~1 together with
  the actual one using iterative gap filling.
\item Calculate the error between the values imputed in step~2
  for artificial gaps and the actual values of the series $\tY$
  in the corresponding interval.
\item Repeat steps~1--3 $R$ times for the given set of $R$
  intervals of artificial missing values. For each interval,
  the error is calculated according to step~3.
\end{enumerate}
\noindent \textit{Result}: The arithmetic average of the $R$
  errors computed in step~4.

\bigskip
In this paper, we use SSA \citep{Golyandina2001,Golyandina2020} as the signal
reconstruction (noise removal) method that is part of the iterative gap-filling
process \citep{Kondrashov2006}.
Let us describe it briefly. The SSA method in its version for signal extraction
has only two parameters, the window length $L$ and the number of components $r$.
At the first stage, the time series $\tX = (x_1,\ldots,x_N)$
is transformed into the so-called trajectory matrix:
\begin{equation}
\label{eq:traj}
\bfX =
\left(\!{\renewcommand{\arraystretch}{1.1}
  \begin{array}{@{\ }l@{\;\;}l@{\;\;}l@{\;\;}l@{\;}}
     x_1       & x_2        & \dots       & x_{K}    \\
     x_2       & \;\ddots   & \;\ddots    & x_{K+1}  \\
     \;\vdots  & \;\ddots   & \ \:\ddots  & \;\vdots \\
     x_{L}     & x_{L+1}    & \dots       & x_{N}
  \end{array}
}\right)
\end{equation}
using the parameter $L$.
Next, the singular value decomposition (SVD) of the trajectory
matrix is constructed.
Let $\bfS=\bfX\bfX^{\rmT}$, $\lambda_1,\ldots,\lambda_L$ be
the eigenvalues of the matrix $\bfS$ taken in non-increasing
order, $U_1,\ldots,U_L$ be the orthonormalized system of
eigenvectors corresponding to these eigenvalues.
Let $d=\max\{k : \lambda_k>0\}$ and $V_k=\bfX^{\rmT}U_k/\sqrt{\lambda_{k}}$, $k=1,\ldots,d$.
Then the SVD of the matrix $\bfX$ can be represented
as a sum of elementary matrices:
\begin{equation}
\label{eq:sumx}
\bfX = \bfX_1+\ldots+\bfX_d,
\end{equation}
where $\bfX_k=\sqrt{\lambda_k}U_k V_k^{\rmT},\ k=1,\ldots,d$.

In the second step, the first $r$ elementary matrices of the SVD are summed:
\begin{equation}
\label{eq:sumy}
\bfY = \bfX_1+\ldots+\bfX_r,
\end{equation}
and their sum is converted into a time series
\begin{equation}
\label{eq:Hank}
\widetilde{y}_{s} = \sum_{(l,k)\in A_s} y_{lk}\Big/|A_s|,
\end{equation}
where $y_{lk}$ are the elements of the matrix $\bfY$, the sets
$A_s=\{(l,k): l+k=s+1, \, 1\!\leqslant\! l\!\leqslant\! L, \, 1\!\leqslant\! k\!\leqslant K\}$, $s=1,\ldots,N$,
give the indices corresponding to the elements on the side
diagonals, $|A_s|$ denotes the number of elements in the sets $A_s$.
The time series $\widetilde\tY=(\widetilde{y}_1,\ldots,\widetilde{y}_N)$
is considered as the SSA estimate of the signal.

\bigskip
The SSA algorithm can be extended to the simultaneous analysis
of two time series, $\tX$ and $\tY$, in two forms.
In the case of Complex SSA (CSSA) for a time series of the
form $x_n + \iu y_n$, $n=1,\ldots,N$, the transpose operator
$\rmT$ is replaced by the Hermite conjugate operator $\rmH$.
The complex form of SSA has been considered since its origin.
In the complex form, the first descriptions of the method can be
found, for example, in \citet{Kumaresan.Tufts1982,Keppenne.Lall1996}.

In the case of Multivariate SSA (MSSA) for two time series of the
$(x_n,y_n)$, $n=1,\ldots,N$, the difference lies in the construction
of the trajectory matrix, which consists of stacked trajectory matrices of the time series.
A more detailed description can be found in \citet{Golyandina.etal2015}.

%%%%%%%%%%%%%%%%%%%%%%%%%%%%%%%%%%%%%%%%

\section*{Acknowledgments}

The authors are grateful to three anonymous reviewers and scientific
editor, Alberto Ecsapa, for valuable comments and suggestions which
helped to improve the manuscript.
The SSA data analysis was performed using the package \texttt{Rssa},
\url{https://CRAN.R-project.org/package=Rssa}.
This research has made use of SAO/NASA Astrophysics Data System (ADS),
\url{https://ui.adsabs.harvard.edu/}.
The figures were prepared using \texttt{gnuplot}, \url{http://www.gnuplot.info/}.

\section*{Author Contributions}

Z.M. designed the paper and performed preliminary data analysis and computations related to the parametric model.
N.G. and R.O. performed the SSA data analysis.
All authors contributed to the writing of the text, discussed the results, and approved the manuscript for submission.

\section*{Data Availability}

IERS C01 EOP series is available at \url{https://datacenter.iers.org/eop.php}
and \url{ftp://hpiers.obspm.fr/iers/eop/eopc01/}.

\section*{Conflicts of interest}

The authors state that there are no known conflicts of interest.

\bibliography{c01_y_gap}

\begin{thebibliography}{49}
\providecommand{\natexlab}[1]{#1}
\providecommand{\doi}[1]{doi:\discretionary{}{}{}#1}
\providecommand{\url}[1]{{#1}}
\providecommand{\eprint}[2][]{\url{#2}}

\bibitem[{{Beutler} et~al.(2020){Beutler}, {Villiger}, {Dach}, {Verdun}, and
  {J{\"a}ggi}}]{Beutler2020AdSpR}
{Beutler} G, {Villiger} A, {Dach} R, {Verdun} A, {J{\"a}ggi} A (2020) {Long
  polar motion series: Facts and insights}. Advances in Space Research
  66(11):2487--2515. \url{https://doi.org/10.1016/j.asr.2020.08.033}

\bibitem[{{Bizouard} et~al.(2019){Bizouard}, {Lambert}, {Gattano}, {Becker},
  and {Richard}}]{Bizouard2019}
{Bizouard} C, {Lambert} S, {Gattano} C, {Becker} O, {Richard} JY (2019) {The
  IERS EOP 14C04 solution for Earth orientation parameters consistent with ITRF
  2014}. Journal of Geodesy 93(5):621--633.
  \url{https://doi.org/10.1007/s00190-018-1186-3}

\bibitem[{{Chandler}(1894{\natexlab{a}})}]{Chandler1894_322}
{Chandler} SC (1894{\natexlab{a}}) {On the inequalities in the coefficients of
  the law of latitude-variation}. \aj 14:73--75.
  \url{https://doi.org/10.1086/102074}

\bibitem[{{Chandler}(1894{\natexlab{b}})}]{Chandler1894_320}
{Chandler} SC (1894{\natexlab{b}}) {Variation of latitude from the Greenwich
  muralcircle Observations, 1836-51}. \aj 14:57--60.
  \url{https://doi.org/10.1086/102062}

\bibitem[{{Dick} et~al.(2000){Dick}, {McCarthy}, and {Luzum}}]{IAUC178}
{Dick} S, {McCarthy} D, {Luzum} B (eds)  (2000) {IAU Colloq. 178: Polar Motion:
  Historical and Scientific Problems}, Astronomical Society of the Pacific
  Conference Series, vol 208. Astronomical Society of the Pacific

\bibitem[{{Elsner} and {Tsonis}(1996)}]{Elsner1996}
{Elsner} JB, {Tsonis} AA (1996) {Singular Spectrum Analysis: A New Tool in Time
  Series Analysis}. Springer

\bibitem[{{Fedorov} et~al.(1972){Fedorov}, {Korsun'}, {Major}, {Panchenko},
  {Taradij}, and {Yatskiv}}]{Fedorov1972}
{Fedorov} EP, {Korsun'} AA, {Major} SP, {Panchenko} NI, {Taradij} VK, {Yatskiv}
  YS (1972) {Motion of the Earth's pole from 1890.0 to 1969.0}. Kyiv, Naukova
  Dumka (in Russian)

\bibitem[{{Gambis}(2000)}]{Gambis2000ASPC}
{Gambis} D (2000) {Long-term Earth Orientation Monitoring Using Various
  Techniques}. In: {Dick} S, {McCarthy} D, {Luzum} B (eds) IAU Colloq. 178:
  Polar Motion: Historical and Scientific Problems, Astronomical Society of the
  Pacific, Astronomical Society of the Pacific Conference Series, vol 208, pp
  337--344

\bibitem[{{Gaposchkin}(1972)}]{Gaposchkin1972}
{Gaposchkin} EM (1972) {Analysis of Pole Position from 1846 to 1970}. In:
  {Melchior} P, {Yumi} S (eds) IAU Symp. 48: Rotation of the Earth, Cambridge
  University Press, vol~48, p 19–32,
  \url{https://doi.org/10.1017/S0074180900098016}

\bibitem[{{Ghil} et~al.(2002){Ghil}, {Allen}, {Dettinger}, {Ide}, {Kondrashov},
  {Mann}, {Robertson}, {Saunders}, {Tian}, {Varadi}, and {Yiou}}]{Ghil2002}
{Ghil} M, {Allen} MR, {Dettinger} MD, {Ide} K, {Kondrashov} D, {Mann} ME,
  {Robertson} AW, {Saunders} A, {Tian} Y, {Varadi} F, {Yiou} P (2002) {Advanced
  Spectral Methods for Climatic Time Series}. Reviews of Geophysics 40(1):1003.
  \url{https://doi.org/10.1029/2000RG000092}

\bibitem[{{Golyandina} and {Osipov}(2007)}]{Golyandina.Osipov2007}
{Golyandina} N, {Osipov} E (2007) The ``{Caterpillar}''--{SSA} method for
  analysis of time series with missing values. J Stat Plan Inference
  137(8):2642--2653. \url{https://doi.org/10.1016/j.jspi.2006.05.014}

\bibitem[{{Golyandina} and {Zhigljavsky}(2020)}]{Golyandina2020}
{Golyandina} N, {Zhigljavsky} A (2020) {Singular Spectrum Analysis for Time
  Series. Second Edition}. Springer,
  \url{https://doi.org/10.1007/978-3-662-62436-4}

\bibitem[{{Golyandina} et~al.(2001){Golyandina}, {Nekrutkin}, and
  {Zhigljavsky}}]{Golyandina2001}
{Golyandina} N, {Nekrutkin} V, {Zhigljavsky} A (2001) {Analysis of Time Series
  Structure: SSA and related techniques}. Chapman and Hall/CRC

\bibitem[{{Golyandina} et~al.(2015){Golyandina}, {Korobeynikov}, {Shlemov}, and
  {Usevich}}]{Golyandina.etal2015}
{Golyandina} N, {Korobeynikov} A, {Shlemov} A, {Usevich} K (2015) Multivariate
  and {2D} Extensions of Singular Spectrum Analysis with the {Rssa} Package. J
  Stat Softw 67(2):1--78. \url{https://doi.org/10.18637/jss.v067.i02}

\bibitem[{{Gross}(1992)}]{Gross1992GJI}
{Gross} RS (1992) {Correspondence between theory and observations of polar
  motion.} Geophysical Journal International 109:162--170.
  \url{https://doi.org/10.1111/j.1365-246X.1992.tb00086.x}

\bibitem[{{Guo} et~al.(2005){Guo}, {Greiner-Mai}, {Ballani}, {Jochmann}, and
  {Shum}}]{Guo2005JGeod}
{Guo} JY, {Greiner-Mai} H, {Ballani} L, {Jochmann} H, {Shum} CK (2005) {On the
  double-peak spectrum of the Chandler wobble}. Journal of Geodesy
  78(11-12):654--659. \url{https://doi.org/10.1007/s00190-004-0431-0}

\bibitem[{{H{\"o}pfner}(2004)}]{Hopfner2004SGeo}
{H{\"o}pfner} J (2004) {Low-Frequency Variations, Chandler and Annual Wobbles
  of Polar Motion as Observed Over One Century}. Surveys in Geophysics
  25(1):1--54. \url{https://doi.org/10.1023/B:GEOP.0000015345.88410.36}

\bibitem[{{Ji} et~al.(2023){Ji}, {Shen}, {Chen}, and {Wang}}]{Ji2023}
{Ji} K, {Shen} Y, {Chen} Q, {Wang} F (2023) {Extended singular spectrum
  analysis for processing incomplete heterogeneous geodetic time series}.
  Journal of Geodesy 97(8):74. \url{https://doi.org/10.1007/s00190-023-01764-8}

\bibitem[{{Keppenne} and {Lall}(1996)}]{Keppenne.Lall1996}
{Keppenne} C, {Lall} U (1996) Complex singular spectrum analysis and
  multivariate adaptive regression splines applied to forecasting the southern
  oscillation. Experimental Long-Lead Bulletin 5(3):39--40

\bibitem[{{Ko{\l}aczek} and {Kosek}(1999)}]{Kolaczek1999srst}
{Ko{\l}aczek} B, {Kosek} W (1999) {Variations of the amplitude of the Chandler
  wobble.} In: Journ\'ees 1998 ``Syst\`emes de R\'ef\'erence Spatio-Temporels:
  Conceptual, Conventional and Practical Studies Related to Earth Rotation'',
  pp 215--220

\bibitem[{{Kondrashov} and {Ghil}(2006)}]{Kondrashov2006}
{Kondrashov} D, {Ghil} M (2006) {Spatio-temporal filling of missing points in
  geophysical data sets}. Nonlinear Processes in Geophysics 13(2):151--159.
  \url{https://doi.org/10.5194/npg-13-151-2006}

\bibitem[{Kumaresan and Tufts(1982)}]{Kumaresan.Tufts1982}
Kumaresan R, Tufts D (1982) Estimating the parameters of exponentially damped
  sinusoids and pole-zero modeling in noise. IEEE Trans Acoust 30(6):833--840

\bibitem[{{Li} et~al.(1974){Li}, {Xu}, and {Lin}}]{LiXuLin1974}
{Li} Z, {Xu} H, {Lin} B (1974) {Coordinates of the earth's instataneous pole
  from 1825.0 to 1897.9.} Acta Astronomica Sinica 15:86--92 (in Chinese)

\bibitem[{{Lopes} et~al.(2022){Lopes}, {Courtillot}, {Gibert}, and {Le
  Mou{\"e}l}}]{Lopes2022Geosc}
{Lopes} F, {Courtillot} V, {Gibert} D, {Le Mou{\"e}l} JL (2022) {Extending the
  Range of Milankovic Cycles and Resulting Global Temperature Variations to
  Shorter Periods (1{\textendash}100 Year Range)}. Geosciences 12(12):448.
  \url{https://doi.org/10.3390/geosciences12120448}

\bibitem[{{Malkin} and {Miller}(2010)}]{Malkin2010}
{Malkin} Z, {Miller} N (2010) {Chandler wobble: two more large phase jumps
  revealed}. Earth, Planets and Space 62(12):943--947.
  \url{https://doi.org/10.5047/eps.2010.11.002}

\bibitem[{{Miller}(2011)}]{Miller2011}
{Miller} NO (2011) {Chandler wobble in variations of the Pulkovo latitude for
  170 years}. Solar System Research 45(4):342--353.
  \url{https://doi.org/10.1134/S0038094611040058}

\bibitem[{{Okhotnikov} and {Golyandina}(2019)}]{Golyandina2019}
{Okhotnikov} G, {Golyandina} N (2019) {EOP} time series prediction using
  singular spectrum analysis. In: {Corpetti} T, {Ienco} D, {Interdonato} R, {et
  al} (eds) Proceedings of {MACLEAN}: {MAChine Learning for EArth ObservatioN
  Workshop}, RWTH Aahen University, Germany, CEUR Workshop Proceedings, Vol.
  2466

\bibitem[{{Orlov}(1961)}]{Orlov1961}
{Orlov} AY (1961) {Free nutation based on observations in Pulkovo in
  1842--1912}. In: {Aksent'eva} ZN (ed) {Orlov A.~Ya., Selected works, Vol.~1},
  Kyiv: Academy of Sciences of USSR, pp 95--113 (in Russian)

\bibitem[{{Petit} and {Luzum}(2010)}]{IERSConv2010}
{Petit} G, {Luzum} B (eds)  (2010) IERS Conventions (2010). IERS Technical Note
  No.~36, Verlag des Bundesamts f\"ur Kartographie und Geod\"asie, Frankfurt am
  Main

\bibitem[{{Rykhlova}(1970{\natexlab{a}})}]{Rykhlova1970SoobGAISH}
{Rykhlova} LV (1970{\natexlab{a}}) {The coordinates of the Earth's pole for the
  years 1846.0-1891.5.} Soobshcheniya Gosudarstvennogo Astronomicheskogo
  Instituta, Moscow State University 163:3--10 (in Russian)

\bibitem[{{Rykhlova}(1970{\natexlab{b}})}]{Rykhlova1970TrGAISH}
{Rykhlova} LV (1970{\natexlab{b}}) {Variations of the Washington latitude in
  1845--1891}. Trudy Gosudarstvennogo Astronomicheskogo Instituta, Moscow State
  University 39:200--212 (in Russian)

\bibitem[{{Rykhlova} and {Kurbasova}(1990)}]{Rykhlova1990IAUS}
{Rykhlova} LV, {Kurbasova} GS (1990) {The Study of the Structure of 142-Year
  Series of Pole Coordinates}. In: {Lieske} JH, {Abalakin} VK (eds) Inertial
  Coordinate System on the Sky, vol 141, p 157

\bibitem[{{Schuh} et~al.(2001){Schuh}, {Nagel}, and {Seitz}}]{Schuh2001JGeod}
{Schuh} H, {Nagel} S, {Seitz} T (2001) {Linear drift and periodic variations
  observed in long time series of polar motion}. Journal of Geodesy
  74(10):701--710. \url{https://doi.org/10.1007/s001900000133}

\bibitem[{{Seitz} et~al.(2012){Seitz}, {Kirschner}, and
  {Neubersch}}]{Seitz2012JGRB}
{Seitz} F, {Kirschner} S, {Neubersch} D (2012) {Determination of the Earth's
  pole tide Love number k$_{2}$ from observations of polar motion using an
  adaptive Kalman filter approach}. Journal of Geophysical Research (Solid
  Earth) 117(B9):B09403. \url{https://doi.org/10.1029/2012JB009296}

\bibitem[{{Sekiguchi}(1975)}]{Sekiguchi1975}
{Sekiguchi} N (1975) {On the latitude variations of the interval between 1830
  and 1860.} Journal of the Geodetic Society of Japan 21:131--141

\bibitem[{{Shen} et~al.(2015){Shen}, {Peng}, and {Li}}]{Shen2015}
{Shen} Y, {Peng} F, {Li} B (2015) {Improved singular spectrum analysis for time
  series with missing data}. Nonlinear Processes in Geophysics 22(4):371--376.
  \url{https://doi.org/10.5194/npg-22-371-201510.5194/npgd-1-1947-2014}

\bibitem[{{Thackeray}(1893)}]{Thackeray1893}
{Thackeray} WG (1893) {Latitude variation and Greenwich observations, 1851-91}.
  \mnras 53:120. \url{https://doi.org/10.1093/mnras/53.3.120}

\bibitem[{{Vautard} and {Ghil}(1989)}]{Vautard1989}
{Vautard} R, {Ghil} M (1989) {Singular spectrum analysis in nonlinear dynamics,
  with applications to paleoclimatic time series}. Physica D Nonlinear
  Phenomena 35:395--424. \url{https://doi.org/10.1016/0167-2789(89)90077-8}

\bibitem[{{Vityazev} et~al.(2010){Vityazev}, {Miller}, and
  {Prudnikova}}]{Vityazev2010AIPC}
{Vityazev} VV, {Miller} NO, {Prudnikova} EJ (2010) {Singular Spectrum Analysis
  in Astrometry and Geodynamics}. In: {de Le{\'o}n} M, {de Diego} DM, {Ros} RM
  (eds) Mathematics and Astronomy: A Joint Long Journey, AIP, American
  Institute of Physics Conference Series, vol 1283, pp 319--328,
  \url{https://doi.org/10.1063/1.3506397}

\bibitem[{{Vondr{\'a}k}(1988)}]{Vondrak1988IAUS}
{Vondr{\'a}k} J (1988) {Is Chandler Frequency Constant?} In: {Babcock} AK,
  {Wilkins} GA (eds) The Earth's Rotation and Reference Frames for Geodesy and
  Geodynamics, IAU Symposium, vol 128, p 359

\bibitem[{{Vondr\'ak} et~al.(1995){Vondr\'ak}, {Ron}, {Pe\u{s}ek}, and
  {\u{C}epek}}]{Vondrak1995}
{Vondr\'ak} J, {Ron} C, {Pe\u{s}ek} I, {\u{C}epek} A (1995) {New global
  solution of Earth orientation parameters from optical astrometry in
  1900-1990}. \aap 297:899--906

\bibitem[{{Vondr\'ak} et~al.(2010){Vondr\'ak}, {Ron}, and
  {\u{S}tefka}}]{Vondrak2010}
{Vondr\'ak} J, {Ron} C, {\u{S}tefka} A (2010) {Earth orientation parameters
  based on EOC-4 astrometric catalog}. Acta Geodyn Geomater 7(3):245–251

\bibitem[{{Yamaguchi} and {Furuya}(2024)}]{Yamaguchi2024}
{Yamaguchi} R, {Furuya} M (2024) {Can we explain the post-2015 absence of the
  Chandler wobble?} Earth, Planets and Space 76(1):1.
  \url{https://doi.org/10.1186/s40623-023-01944-y}

\bibitem[{{Yatskiv}(2000)}]{Yatskiv2000ASPC}
{Yatskiv} Y (2000) {Chandler Motion Observations}. In: {Dick} S, {McCarthy} D,
  {Luzum} B (eds) IAU Colloq. 178: Polar Motion: Historical and Scientific
  Problems, Astronomical Society of the Pacific, Astronomical Society of the
  Pacific Conference Series, vol 208, pp 383--395

\bibitem[{{Yi} and {Sneeuw}(2021)}]{Yi2021}
{Yi} S, {Sneeuw} N (2021) {Filling the Data Gaps Within GRACE Missions Using
  Singular Spectrum Analysis}. Journal of Geophysical Research (Solid Earth)
  126(5):e2020JB021227. \url{https://doi.org/10.1029/2020JB021227}

\bibitem[{{Zechmeister} and {K{\"u}rster}(2009)}]{Zechmeister2009}
{Zechmeister} M, {K{\"u}rster} M (2009) {The generalised Lomb-Scargle
  periodogram. A new formalism for the floating-mean and Keplerian
  periodograms}. \aap 496(2):577--584.
  \url{https://doi.org/10.1051/0004-6361:200811296}

\bibitem[{{Zotov} and {Bizouard}(2012)}]{Zotov2012JGeo}
{Zotov} L, {Bizouard} C (2012) {On modulations of the Chandler wobble
  excitation}. Journal of Geodynamics 62:30--34.
  \url{https://doi.org/10.1016/j.jog.2012.03.010}

\bibitem[{{Zotov}(2010)}]{Zotov2010ArtSa}
{Zotov} LV (2010) {Dynamical Modeling and Excitation Reconstruction as
  Fundamental of Earth Rotation Prediction}. Artificial Satellites
  45(2):95--106. \url{https://doi.org/10.2478/v10018-010-0010-y}

\bibitem[{{Zotov} et~al.(2022){Zotov}, {Sidorenkov}, and
  {Bizouard}}]{Zotov2022MUPB}
{Zotov} LV, {Sidorenkov} NS, {Bizouard} C (2022) {Anomalies of the Chandler
  Wobble in 2010s}. Moscow University Physics Bulletin 77(3):555--563.
  \url{https://doi.org/10.3103/S0027134922030134}

\end{thebibliography}
\bibliographystyle{joge}
\end{document}